\documentclass[11pt]{article}
\usepackage{graphicx}
\usepackage{cite}

\newcommand{\BABARPubYear}    {02}

\newcommand{\BABARConfNumber} {10}
\newcommand{\SLACPubNumber} {9315}

\input pubboard/babarsym
\def\br{{\cal B}}
\def\bdsta1{B^0 \rightarrow D^{*-} a_1^+}
\def\bnonres{B^0 \rightarrow D^{*-} \rho^0 \pi^+}
\def\btot{B^0 \rightarrow D^{*-} \pi^+ \pi^- \pi^+}
\def\bddsta1{B \rightarrow D^{**} a_1}
\def\ddtodst{D^{**} \rightarrow D^{*} \pi}

\def\dstminus{D^{*-}}
\def\dzbar{\overline D^0}
\def\sincp{\sin(2\beta + \gamma)}
\def\mmiss{m_{\rm miss}}
\def\ups{\Upsilon(4S)}
\def\a1Helic{\cos\theta_{a1^+}^H}

\def\dsta1{D^{*} a_1}

\long\def\inst#1{\par\nobreak\kern 4pt\nobreak
    {\it #1}\par\vskip 10pt plus 3pt minus 3pt}

\begin{document}
{\pagestyle{empty}

\begin{flushright}
\babar-CONF-\BABARPubYear/\BABARConfNumber \\
SLAC-PUB-\SLACPubNumber \\
July 2002 \\
\end{flushright}

\par\vskip 5cm

\begin{center}
\Large {\bf Measurement of the {\boldmath $\bdsta1$} Branching Fraction
            with Partially Reconstructed {\boldmath $D^*$} }
\end{center}
\bigskip

\begin{center}
\large The \babar\ Collaboration\\
\mbox{ }\\
July 25, 2002
\end{center}
\bigskip \bigskip

\begin{center}
\large \bf Abstract
\end{center}

The $\bdsta1$ branching fraction has been measured with data collected by the \babar\ experiment in
1999 and 2000 corresponding to a total integrated luminosity of $20.6~{\rm fb}^{-1}$. Signal events 
have been selected using a partial reconstruction technique, in which only the $a_1^+$ and the slow 
pion ($\pi_s$) from the $\dstminus$ decay are identified. A signal yield of $18400 \pm 1200$ events 
has been found, corresponding to a preliminary branching fraction of $(1.20 \pm 0.07$(stat) $\pm 0.14$(syst)$)\%$.

\vfill
\begin{center}
Contributed to the 31$^{st}$ International Conference on High Energy Physics,\\
7/24---7/31/2002, Amsterdam, The Netherlands
\end{center}

\vspace{1.0cm}
\begin{center}
{\em Stanford Linear Accelerator Center, Stanford University, 
Stanford, CA 94309} \\ \vspace{0.1cm}\hrule\vspace{0.1cm}
Work supported in part by Department of Energy contract DE-AC03-76SF00515.
\end{center}

\newpage
} 

\begin{center}
\small

The \babar\ Collaboration,
\bigskip

B.~Aubert,
D.~Boutigny,
J.-M.~Gaillard,
A.~Hicheur,
Y.~Karyotakis,
J.~P.~Lees,
P.~Robbe,
V.~Tisserand,
A.~Zghiche
\inst{Laboratoire de Physique des Particules, F-74941 Annecy-le-Vieux, France }
A.~Palano,
A.~Pompili
\inst{Universit\`a di Bari, Dipartimento di Fisica and INFN, I-70126 Bari, Italy }
% ...

\end{center}\newpage

\setcounter{footnote}{0}

\section{Introduction}
\label{sec:intro}

A partial reconstruction technique has been used in the past to select large samples of reconstructed $B$ 
mesons with a $D^{*-}$ in the final state~\cite{ref:cleo} and to measure properties of the $B^0$ meson.
In this method the $\dzbar$ is not reconstructed, but its four-momentum is inferred from the kinematics 
of the $a_1^+$, the slow pion ($\pi_s$) from $D^{*-}$ decay and the decay constraints. The measurement
of the branching fraction for $\bdsta1$ is performed as a first step in
demonstrating that the partial reconstruction method may offer a means of determining the combination of
Cabibbo-Kobayashi-Maskawa\cite{ref:ckm} unitarity triangle angles $\sincp$ using this 
channel\footnote{Since the selection of $D^{*-}a_1^+$ and $D^{*+}a_1^-$ are identical, 
the charge conjugate state is implied throughout the paper.}. 


\section {The \babar\ detector}
\label{sec:detector}

The \babar\ experiment is located at the PEP-II storage ring at the Stanford Linear Accelerator 
Center. A detailed description of the detector and of the algorithms used for the track 
reconstruction and selection of $B\overline{B}$ events can be found in Ref.~\cite{ref:babardet}. 
For the partial reconstruction analysis of $\bdsta1$ only charged tracks are used: particles 
with transverse momentum $p_T > 170$~MeV/c are reconstructed by matching hits in the Silicon Vertex Tracker 
(SVT) with track segments in the Drift Chamber (DCH). Tracks with lower $p_T$ do not penetrate a significant
distance in the DCH and are reconstructed using only the information from the SVT.

Electron, muon and kaon identification is used in the analysis as a veto in the selection of 
pions forming the $a_1$ candidates. Electron candidates are identified by the ratio 
of the energy deposited in the electromagnetic calorimeter (EMC) to the track momentum ($E/p$) 
and by the energy loss in the DCH ($dE/dx$). Muons are primarily identified by the measured number 
of hadronic interaction lengths traversed from the outside radius of the DCH through the iron of the
instrumented flux return (IFR). Kaons are distinguished from pions and protons on the basis 
of $dE/dx$ in the SVT and DCH, and the number of Cherenkov photons and the Cherenkov angle in 
the Detector for Internally Reflected Cherenkov radiation (DIRC).


\section{Data sample}
\label{sec:datasample}

The data used in this analysis were collected with the \babar\ detector in 1999 and 2000.
These data correspond to an integrated luminosity of $20.6~{\rm fb}^{-1}$ recorded at the $\ups$ 
resonance and, for background studies, $2.6~{\rm fb}^{-1}$ collected at 40 MeV below the resonance 
(``off-resonance'' sample). Monte Carlo samples of $B\overline{B}$ and continuum events 
were reconstructed and analyzed using the same procedure as the data. The equivalent luminosity of the 
generic simulated data is approximately one fourth of the on-resonance data, while the number 
of signal $\bdsta1$ Monte Carlo (MC) events is approximately ten times the number expected in 
the on-resonance data.


\section{The partial reconstruction technique}
\label{sec:partrec}

In the partial reconstruction of the decay chain $\bdsta1$, followed by $D^{*-} \rightarrow {\overline D^0}
\pi_s^-$, only the $a_1$ and the $\pi_s$ from $D^*$ decay are required. For this analysis 
the $a_1$ is reconstructed via the decay chain $a_1^+ \rightarrow \rho^0 \pi^+$. 
The angle between the momentum vectors of the $B$ and the $a_1$ in the center-of-mass frame (CM) is then computed:

\begin{equation}
\cos \theta_{Ba_1} = \frac{M^2_{D^{*-}} - M^2_{B^0} - M^2_{a_1} + E_{CM} E_{a_1}}
                          {2P_{B}|\vec{p}_{a_1}|}
\label{eqn:ba1angle}
\end{equation}

\noindent
where $M_x$ is the mass of particle x, $E_{a_1}$ and $\vec{p}_{a_1}$ are the measured CM energy and 
momentum of the $a_1$, $E_{CM}$ is the total CM energy of the beams and 
$P_{B} = \sqrt{E_{CM}^2/4 - M_{B^0}^2}$. Given $\cos \theta_{Ba_1}$ and the measured four-momenta 
of the $\pi_s$ and the $a_1$, the $B$ four-momentum can be calculated up to an unknown azimuthal 
angle $\phi$ around $\vec{p}_{a_1}$. For every value of $\phi$, the expected ${\overline D^0}$ 
four-momentum, $P_{\overline D^0}$, is determined from four-momentum conservation and the 
$\phi$-dependent ``missing mass'' is calculated, $\mmiss(\phi) \equiv \sqrt{|P_{\overline D^0}|^2}$. 
Defining $m_{min}$ and $m_{max}$ to be the minimum and maximum values of $\mmiss(\phi)$ obtained by
varying $\phi$ over the allowed range, the missing mass is calculated as 
$\mmiss \equiv \frac{1}{2} (m_{max} + m_{min})$.
For signal events, this variable peaks at the ${\overline D^0}$ mass, while for background events 
it has a broader distribution. For this reason, $\mmiss$ can be used to determine the fractions 
of signal and background events in the data sample.


\section{Event selection}
\label{sec:evtsel}

Data and Monte Carlo events are first selected with the following loose requirements:

\begin{itemize}
 \item[$\cdot$] $R2$, the ratio of the 2nd to the 0th Fox Wolfram moments~\cite{ref:book},
                less than 0.35;
 \item[$\cdot$] at least one $a_1$ candidate\footnote{By $a_{1}^{+}$ we refer to the $\pi^+~\pi^-~\pi^+$
                final state.}  such 
that:
  \begin{itemize}
   \item[-] the $a_1$ invariant mass $m_{a_1}$ is between $1.0$ and $1.6$~GeV/c$^2$;
   \item[-] the $a_1$ momentum $p_{a_1}$, computed in the CM frame, is between $1.85$ and $2.30$~GeV/c; 
   \item[-] the vertex probability obtained from a vertex fit of the 3 pions is greater than 1\%;
   \item[-] the invariant mass of at least one of the two possible $\pi^+ \pi^-$ combinations ($m_{\pi \pi}$) 
            is in the range [0.278, 1.122]~GeV/c$^2$;
  \end{itemize}
 \item[$\cdot$] at least one additional track (the $\pi_s$) with CM momentum $p_{\pi_s}$ between $50$ and $700$ MeV/c;
 \item[$\cdot$] for each selected $\dsta1$ candidate, there must be at least 2 additional particles (charged or neutral).
\end{itemize}

\noindent
The fraction of events selected by applying these cuts is summarized in Table ~\ref{tab:eff}
for signal and background MC samples and for off-resonance data.

The main source of background in this analysis is continuum $q\overline{q}$ events, where $q=u,d,s,c$. 
A neural network is used to optimize the separation of $B\overline{B}$ events from the continuum background,
independent of any particular $B$ decay channel. The NN has three layers, with 11 input nodes, 15 hidden nodes 
and one output. Its definition relies on the different topologies of signal and background at the $\FourS$: 
while $B\overline{B}$ events are more ``spherical'', $q\overline{q}$ events are more ``jet-like''. 
The variables used to discriminate jet-like from isotropic events are $R_2$; the thrust of the event~\cite{ref:book}; 
the two invariant masses squared, obtained by adding the four momenta of all particles going into each of the 
hemispheres divided by the plane perpendicular to the thrust axis; and the angle of the thrust axis with respect
to the $e^+e^-$ direction.

The signal to background separation becomes more complicated when one or more gluons are emitted, 
which is more likely to happen when light quarks are produced. In this case there are multiple 
preferred axes, so that the event shape more closely resembles that of the signal. To 
discriminate further between signal and background in this case, the tracks are clustered into 3 
or 4 jets using the so called ``Durham'' algorithm~\cite{ref:durham, ref:durhamws}. 

The event is first clustered into four jets and the following discriminating variables are added to those 
previously defined in order to build up the network: ${\mathbf y_{4}}$, the jet metric~\cite{ref:durham, ref:durhamws} 
obtained when the event is clustered in 4 jets; the QCD matrix element~\cite{ref:aleph1}; the cosine of the maximum 
angle between each pair of jets; the angle between the plane defined by the two most energetic jets and that defined 
by the other two jets; and the angle between the lowest and second lowest energy jet. The event is then further
clustered into 3 jets and the jet metric ${\mathbf y_{3}}$ is also added to the list of network input variables.

\begin{figure}[htbp]
\begin{center}
\includegraphics[height=15cm]{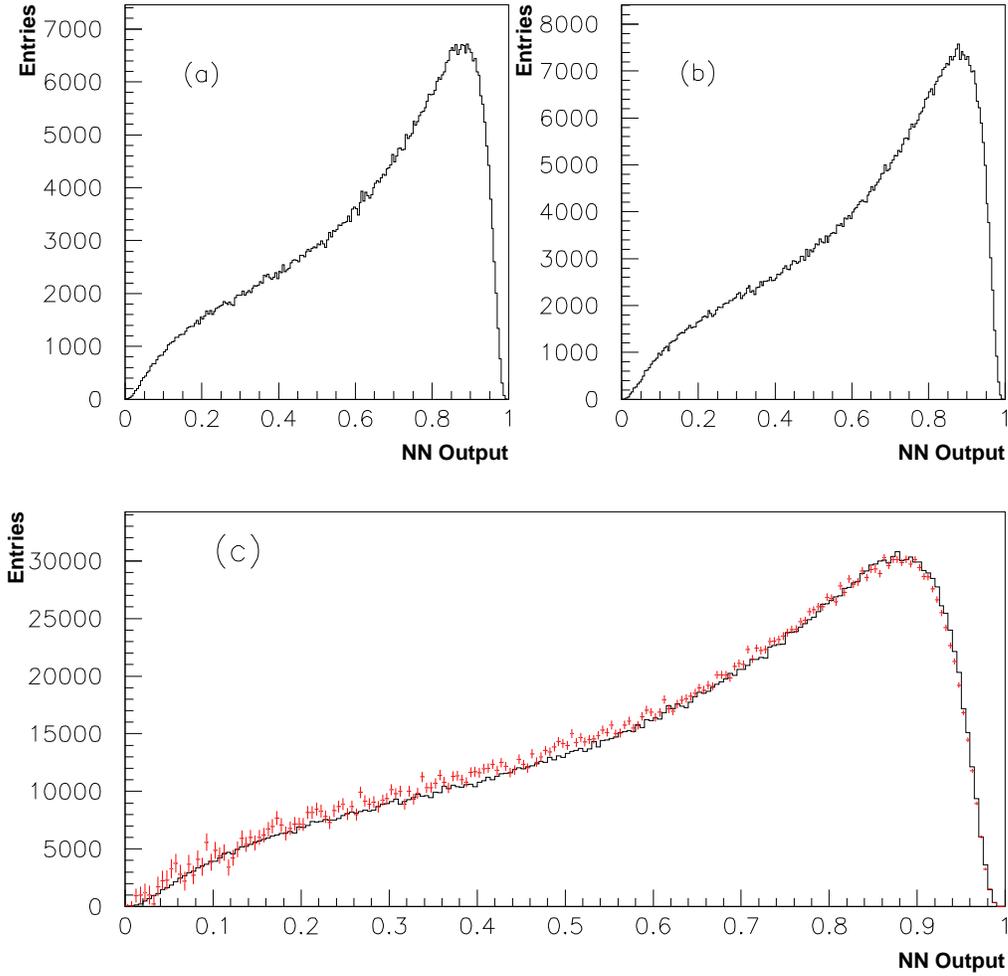}
\caption{  Distribution of the neural net output for (a) $B^0\overline{B^0}$ MC,
	   (b) $B^+B^-$ MC and (c) on-resonance (off-resonance subtracted) data events,
           with superimposed distribution for all $B\overline{B}$ MC events. All events satisfy the 
           event selection criteria. }
\label{fig:outannbbbar}
\end{center}
\end{figure}

The distributions of $B\overline{B}$ MC events and on-resonance data that pass the event selection criteria are shown
in Fig.~\ref{fig:outannbbbar}. The on-resonance distribution is shown after the subtraction of the off-resonance
distribution, scaled by the ratio of the on-resonance to off-resonance luminosities and the CM energy squared. 
By selecting events for which the NN output ($O_{NN}$) is greater than 0.25, 66\% of the continuum events are rejected.
The fractions of events selected by this requirement in the various analyzed samples are shown in Table~\ref{tab:eff}.

After the NN cut, further selection requirements are applied. There must be at least one $a_1$$\pi$ 
combination with net charge equal 0; the three pion invariant mass must be in the range [1.0, 1.45]~GeV/c$^2$; 
the three pion momentum in the CM frame must be in the range [1.9, 2.25]~GeV/c; at least one $\pi^+ \pi^-$ 
combination from the three daughters of the $a_1$ must have invariant mass in the range [0.65, 0.9]~GeV/c$^2$; 
and the vertex probability of the 3$\pi$ plus the $\pi_s$ must be greater than 3\%.

The fractions of events selected by these requirements are summarized in Table~\ref{tab:eff}.

\begin{table}[htb]
\caption{Fractions of events in MC data that are selected by the requirements applied in the analysis.}
\begin{center}
\begin{tabular}{|c|c|c|c|c|}
\hline
Cuts applied & $\bdsta1$ & $B^0 \overline{B^0}$& $B^+ B^-$  & $q\overline{q}$ \\
             &           &    (no signal)      &            &      MC         \\
\hline\hline
Reconstruction Efficiency                 & 38.3\% &  14.6\%  &     13.9\% &  6.4\%  \\
\hline
$O_{NN} > 0.25$                           & 85.4\% &  91.9\%  &     92.1\% & 45.2\%  \\
$q_{a_1-\pi} = 0$                         & 94.2\% &  86.7\%  &     85.2\% & 83.5\%  \\
$1.0 \leq m_{a_1}\leq 1.45$ GeV/$c^2$     & 81.7\% &  71.8\%  &     71.7\% & 73.6\%  \\
$1.9 \leq p_{a_1}\leq 2.25$ GeV/c         & 94.8\% &  92.4\%  &     92.5\% & 93.3\%  \\
$0.65 \leq m_{\pi \pi}\leq 0.9$ GeV/$c^2$ & 76.4\% &  59.1\%  &     60.5\% & 64.3\%  \\
Vertex Prob $4\pi > 0.03$                 & 80.0\% &  74.2\%  &     73.9\% & 76.9\%  \\
\hline\hline
Total Efficiency                          & 14.6\% &   3.4\%  &      3.2\% & 0.82\%  \\
\hline
\end{tabular}
\end{center}
\label{tab:eff}
\end{table}


\section{Results}
\label{sec:results}

Applying all of the selection criteria, the $\mmiss$ distribution of on-resonance and 
off-resonance data is obtained for ``right-sign''events, which are events where the $a_1$ 
and $\pi_s$ candidates have opposite electrical charges. The off-resonance distribution 
is then scaled to take into account the difference in luminosities and CM energies between 
the two samples and subtracted bin-by-bin from the on-resonance distribution. The resulting 
plot is fitted, using a minimum $\chi^2$ technique, to a linear combination of the $\mmiss$ 
distributions of:

\begin{enumerate}
\item $B\overline{B}$ Monte Carlo events, excluding correctly reconstructed signal events; and
\item correctly reconstructed signal Monte Carlo events.
\end{enumerate}

\begin{figure}[htbp]
\begin{center}
\includegraphics[height=15cm]{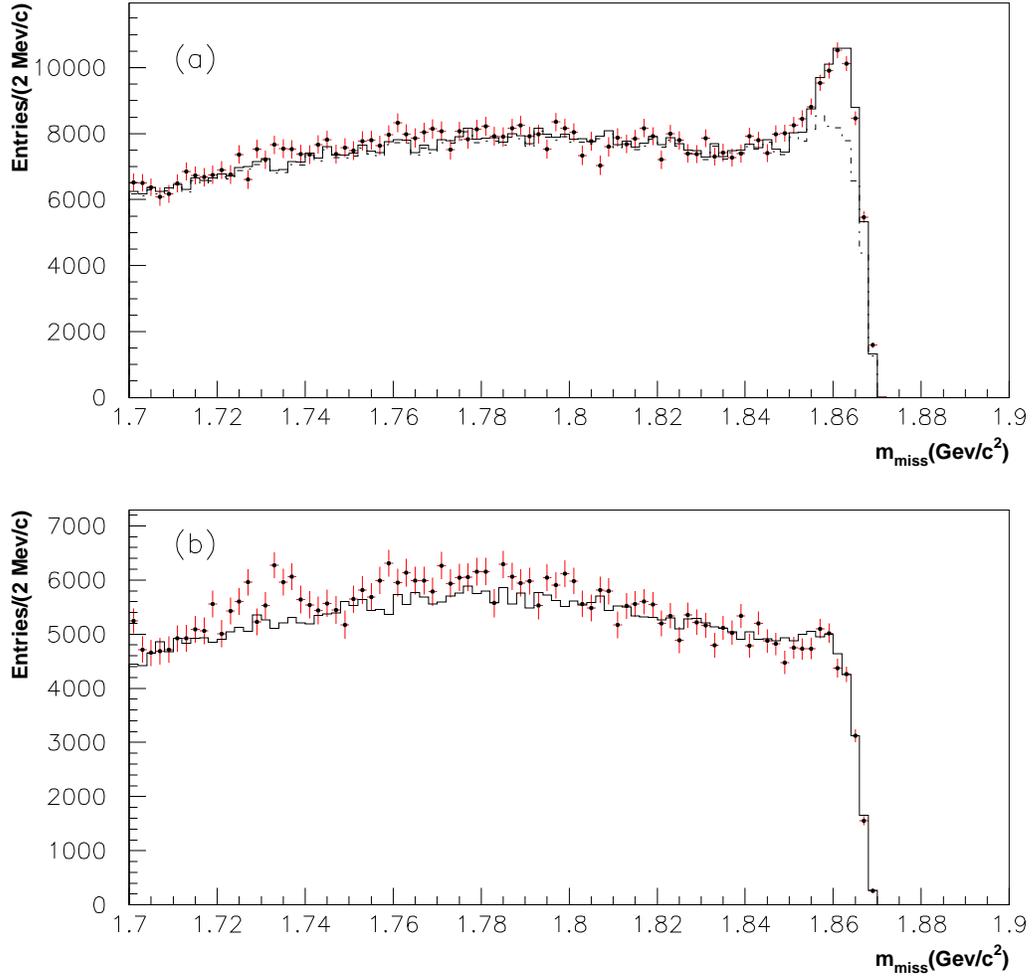}
\caption{ (a) $\mmiss$ distribution of continuum-subtracted on-resonance data events (data points),
          $B\overline{B}$ background MC events (dashed histogram) and $B\overline{B}$ background plus
          signal events (solid histogram) for ``right-sign'' $a_1 \pi$ combinations. The histograms 
          are the result of the fit procedure described in the text.
          (b) Same distributions for ``wrong-sign'' $a_1 \pi$ combinations.}
\label{fig:mmiss-fix-cont}
\end{center}
\end{figure}

In Fig.~\ref{fig:mmiss-fix-cont}(a) the $\mmiss$ distribution of ``right-sign'' 
on-resonance data, off-resonance subtracted, is shown, together with the distributions of
$B\overline{B}$ background MC and signal MC events. The signal yield that is obtained 
is $18400 \pm 1200$ events. The $B\overline{B}$ contribution to the fit is $0.995 \pm 0.015$ of the 
value expected, given the total number of $B\overline{B}$ events in the data and in the Monte Carlo
simulation.

Given this yield, the $a_1$ branching fraction\footnote{In this analysis it is assumed that 
$\br(a_1^+ \rightarrow \rho^0 \pi^+) = 0.4920$, based on isospin considerations and phase space 
corrections.}, the total signal efficiency of Table~\ref{tab:eff} and the fraction of events with
multiple signal candidates, the following preliminary result for the $\bdsta1$ branching fraction is 
obtained:
\begin{equation}
\br(\bdsta1) = (1.20 \pm 0.07)\%,
\label{eq:br}
\end{equation}
\noindent
where the error is statistical only. The result is in very good agreement with the current best 
measurement of ($1.30\pm 0.27$)\%~\cite{ref:pdg,ref:argus2,ref:cleo-br1,ref:cleo-br2}.

Several tests were conducted to verify that the background shapes in data and MC agree and 
do not give rise to a spurious signal. Charged $a_1$ candidates were combined with tracks of the same 
charge into ``wrong-sign'' combinations, and were analyzed in the same way as ``right-sign'' 
combinations. The $\mmiss$ distribution of these candidates, shown in Fig.~\ref{fig:mmiss-fix-cont}(b),
shows a good agreement between data and MC in the signal region ($\mmiss > 1.854$~GeV/c$^2$). 
The $B\overline{B}$ contribution to the fit is $0.990 \pm 0.016$ of the value expected from MC simulation. 

Among events collected on-resonance, the fraction containing more than one $a_1 \pi$ combination
that passes all the analysis requirements in the signal region, defined by $\mmiss > 1.854$~GeV/c$^2$, is
$F = (15.10 \pm 0.14)$\%. This agrees with the fraction $F_{MC} = (15.51 \pm 0.26)$\% obtained from
a weighted mix of off-resonance, $B \overline{B}$ and signal MC events. It was verified that the 
reconstruction efficiency in $q\overline{q}$ MC and off-resonance data are in good agreement.


\section{Systematic errors}
\label{sec:syst}

The selection criteria applied in the analysis are varied within a
reasonable range around their chosen values. The
branching fraction and its error are recalculated for each value, obtaining
$N$ different measurements for $N$ different choices of
the requirement~\cite{ref:aleph}.
The variation of the $N$ results with respect to their average, taking
into account statistical correlations between them~\cite{ref:glen}, is
taken as a systematic error. The method of Ref.~\cite{ref:pdg2} is used
in order to disentangle the statistical fluctuations from the
systematic ones.

The final contributions to the error on the branching ratio due to the variation of the selection
requirements are shown in Table~\ref{tab:syst}. 
It has been verified that the systematic error obtained for each requirement does not depend on the 
variation studied.

\begin{table}[htb]
\caption{Systematic errors on $\br(\bdsta1)$ due to changing the value of some of
         the requirements in the analysis. }
\begin{center}
\begin{tabular}{|c|l|c|c|c|}
\hline
Cut applied                                & Range Studied               & $\tilde{\chi}^2_{min}$ & N &  Error (\%) \\
\hline
$O_{NN} > O_{\rm min}$                     & $O_{\rm min}=$ 0.15 to 0.35 &           1.37         & 5 &   4.2       \\
$m_{\rm min}\leq m_{a_1} \leq m_{\rm max}$ & $m_{\rm min}=$ 1.0~GeV/$c^2$&                        &   &             \\
                                           & $m_{\rm max}=$ 1.4 to 1.5~GeV/$c^2$ &   0.99         & 5 &   negl.     \\
$p_{\rm min}\leq p_{a_1} \leq p_{\rm max}$ & $p_{\rm min}=$ 1.85 to 1.95~GeV/$c^2$ &              &   &             \\
                                           & $p_{\rm max}=$ 2.15 to 2.25~GeV/$c^2$ & 1.96         & 5 &   6.6       \\
Vertex Prob $3\pi > P_{\rm min}$           & $P_{\rm min}=$ 0.01 to 0.15&            1.15         & 5 &   2.4       \\
Vertex Prob $4\pi > P_{\rm min}$           & $P_{\rm min}=$ 0.01 to 0.1 &            1.10         & 5 &   2.2       \\
$m_{\rm min}\leq m_{\rho}\leq m_{\rm max}$ & $m_{\rm min}=$ 0.65 to 0.75~GeV/$c^2$ &              &   &             \\
                                           & $m_{\rm max}=$ 0.8 to 0.9GeV/$c^2$    & 1.45         & 5 &   4.6       \\
\hline
Total error                                &                             &                        &   &   9.6       \\
\hline
\end{tabular}
\end{center}
\label{tab:syst}
\end{table}

A conservative systematic error of 0.35\% is determined from MC due to the dependence of the $\pi_s$ reconstruction
efficiency on the $\pi_s$ momentum. The systematic error due to track reconstruction efficiency is 4.2\% and the 
uncertainty in the total number of $B$ mesons in the data sample is 1.6\%. 

The $B\overline{B}$ background has a component that peaks slightly under the signal region
(Fig.~\ref{fig:mmiss-fix-cont}).
This is due mostly to signal events in which one or more of the selected tracks did not originate from the
signal $B^0$. The contribution of this background is varied in the $B\overline{B}$ MC by 
$\pm \sqrt{(0.07/1.2)^2+0.104^2}$, i.e., by the relative statistical error in the 
central value of the branching fraction plus the total relative systematic error calculated up to 
this point, added in quadrature. This results in a 4.5\% variation of the signal yield, which is 
added to the total systematic error. The total systematic error is 11.5\%. 

Other possible sources of systematic error have been investigated. The non-resonant decay channel $\bnonres$ 
could contribute to the quoted branching fraction. The measured branching fraction for this mode
is $0.57 \pm 0.31\%$~\cite{ref:pdg, ref:cleo-br1}. Since the central value is inconsistent with the total $\br(\btot)$,
this channel is ignored in our fit. As a result, there is a potential contribution from this channel that is 
included in the quoted branching fraction for $\bdsta1$. The central value will shift according to the branching
fraction for $\bnonres$ at a rate of $-3.3\% \times \br(\bnonres)/0.57\%$.

Likewise, the decay $\bddsta1$ could affect the signal yield\footnote{$D^{**}$ denotes
the sum of $D_1$, $D_1^{'}$ and $D_2^{*}$ states.}. To study its effect, $D^{**}$ MC events are added to the generic 
$B\overline{B}$ sample at the level of $\br^{**} = \br(\bddsta1) \times \br(\ddtodst) = 0.35\%$~\cite{ref:HN}, 
and the fit to the missing mass distribution is repeated. This results in a reduction of the signal yield of 4.3\%. 
Since the $\br(\bddsta1)$ has not yet been measured, based on this result the branching fraction obtained in this
analysis will be shifted by contributions from $\bddsta1$ at a rate of $-4.3\% \times \br^{**}/0.35\%$.


\section{Conclusions}
\label{sec:conclusions}

With a partial reconstruction technique, $18400 \pm 1200$ $\bdsta1$ events have
been found in the \babar\ data set of $20.6~{\rm fb}^{-1}$ on-resonance 
events. This corresponds to the branching fraction
\begin{equation}
\br(\bdsta1) = (1.20 \pm 0.07 \pm 0.14)\%,
\end{equation}

\noindent
where the first error is statistical and the second is systematic. Due to the uncertainty in
the contribution of $\bnonres$ events in the signal sample, the central value of the quoted 
branching fraction will shift according to the branching fraction for $\bnonres$ at a rate of
$-3.3\% \times \br(\bnonres)/0.57\%$. Likewise, due to the unknown value of 
$\br^{**} = \br(\bddsta1) \times \br(\ddtodst)$ the central value of the branching fraction will be
shifted by contributions from $\bddsta1$ events at a rate of $-4.3\% \times \br^{**}/0.35\%$. 
The result is in good agreement with the current world average value 
of ($1.30\pm 0.27$)\%~\cite{ref:pdg,ref:argus2,ref:cleo-br1,ref:cleo-br2} but reduces the uncertainty
by a factor of two.
%

\section{Acknowledgments}
\label{sec:Acknowledgments}
We are grateful for the 
extraordinary contributions of our \pep2\ colleagues in
achieving the excellent luminosity and machine conditions
that have made this work possible.
The success of this project also relies critically on the 
expertise and dedication of the computing organizations that 
support \babar.
The collaborating institutions wish to thank 
SLAC for its support and the kind hospitality extended to them. 
This work is supported by the
US Department of Energy
and National Science Foundation, the
Natural Sciences and Engineering Research Council (Canada),
Institute of High Energy Physics (China), the
Commissariat \`a l'Energie Atomique and
Institut National de Physique Nucl\'eaire et de Physique des Particules
(France), the
Bundesministerium f\"ur Bildung und Forschung and
Deutsche Forschungsgemeinschaft
(Germany), the
Istituto Nazionale di Fisica Nucleare (Italy),
the Research Council of Norway, the
Ministry of Science and Technology of the Russian Federation, and the
Particle Physics and Astronomy Research Council (United Kingdom). 
Individuals have received support from 
the A. P. Sloan Foundation, 
the Research Corporation,
and the Alexander von Humboldt Foundation.

\end{document}